\documentclass[final,twocolumn]{elsarticle}

\usepackage{amsmath, amssymb}
\usepackage{bm}
\usepackage{graphicx}
\usepackage[caption=false]{subfig}
\usepackage{mathtools}
\usepackage[breaklinks,colorlinks]{hyperref}
\usepackage{placeins}

\hypersetup{%
,urlcolor=blue
,citecolor=blue
,linkcolor=blue
}

\begin{document}

\title{Vortex stability in pseudo-Hermitian theories}

\author[manchester]{R. A. Battye}
\author[manchester]{S. J. Cotterill}
\author[manchester]{P. Millington}

\address[manchester]{Department of Physics and Astronomy, University of Manchester, Manchester M13 9PL, UK}

\begin{abstract}
Pseudo-Hermitian (including $\mathcal{PT}$-symmetric) field theories support phenomenology that cannot be replicated in standard Hermitian theories.  We describe a concrete example in which the vortex solutions that are realised in a prototypical pseudo-Hermitian field theory exhibit a novel metastability, despite the model parameters residing within the naively stable regime of exact antilinear symmetry of the vacuum theory. This instability is identified analytically and confirmed through numerical simulations, and it arises from the small breaking of the underlying antilinear symmetry of the pseudo-Hermitian theory due to the presence of the topological defect. This leads to spacetime-dependent parameters in the equations of motion governing fluctuations around the vortex, inducing non-trivial exceptional-point structures and complex frequencies within their spectrum. Aside from offering intriguing possibilities for cosmology, this result serves to illustrate the ability to produce long-lived metastable configurations in pseudo-Hermitian field theories of relevance beyond cosmology and high energy physics.
\end{abstract}

\maketitle

\section{Introduction}

When model building new field theories, we often apply various restrictions, for example, imposing symmetries that constrain the terms that can be written down in the Lagrangian or the resulting equations of motion. One of these restrictions is Hermiticity of the Lagrangian or, equivalently, the Hamiltonian. This guarantees that the mass spectrum of fluctuations around a global minimum of the potential will be real. 
However, it is now well established that Hermiticity is not a necessary condition \cite{PhysRevLett.80.5243,10.1063/1.1418246,Mannheim_2018} with several examples of so-called pseudo-Hermitian or $\mathcal{PT}$-symmetric models appearing in the literature. Unlike their Hermitian counterparts, pseudo-Hermitian theories exhibit two regimes of behaviour. One is characterised by a real eigenspectrum owing to the presence of an exact antilinear symmetry of the Hamiltonian, and a common example is exact symmetry under the combined action of parity $\mathcal{P}$ and time-reversal $\mathcal{T}$. The other regime is characterised by a spectrum that contains complex-conjugate pairs of eigenfrequencies. This is referred to as the regime of broken antilinear symmetry, and the boundary between broken and unbroken regimes is marked by so-called exceptional points, at which degrees of freedom are lost.

Given the success of pseudo-Hermitian quantum mechanics across a range of disciplines (see, e.g., Refs.~\cite{ElGanainy:2018ksn, Ashida:2020dkc, Wang:23}), there is increasing interest in pseudo-Hermitian quantum field theories and, more recently, in non-linear theories, where one might anticipate a richer phenomenology. Such non-linear pseudo-Hermitian theories can exhibit topological defects~\cite{Begun:2022ufc,CORREA2021115516,CORREA2022115783,Fring_2020,FRING2020135583}, and these extended field configurations provide an interesting and non-trivial probe of non-Hermitian physics.

In particular, it is well known that pseudo-Hermitian theories with a real eigenspectrum can be related to a Hermitian theory via a similarity transformation. However, for models with non-linear dynamics such as the one considered here, this similarity transformation can become local. It is therefore possible for parts of the spectrum to have passed through the exceptional point, such that we cannot map onto a Hermitian theory. Notably, this means that although the linearised dynamics of the system around a stable, homogeneous vacuum state in the theory can be equivalently described by mapping to a Hermitian system, the inhomogeneous, full dynamics of the system cannot be.

In this {\em letter}, we undertake numerical simulations of the vortex solutions first analysed in Ref.~\cite{Begun:2022ufc}. We find that these solutions can be metastable, even in the would-be unbroken regime of the antilinear symmetry, having an instability that appears on comparatively long time scales. We identify that the origin of this instability is the explicit spacetime dependence of the equations of motion governing fluctuations about the vortex solutions, confirming earlier expectations that the spectra of pseudo-Hermitian field theories with local Lagrangian parameters should exhibit momentum-dependent exceptional points~\cite{Chernodub:2021waz,Chernodub:2024lkr}, such that some modes lie in the broken regime and others in the unbroken regime of the antilinear symmetry. We are able to identify the eigenmode most sensitive to this spontaneous breaking of the exact antilinear symmetry of the theory, and the imaginary part of its eigenfrequency is the dominant driver of the metastability. These semi-analytic results are corroborated by the numerical simulations.

These results confirm that inhomogeneous solutions to the field equations of pseudo-Hermitian theories can be inherently unstable in certain regimes of parameter space that would otherwise be expected to be stable based on an analysis of only homogeneous configurations. This represents an unusual dynamical behaviour for topological defect solutions and may find applications from cosmology and high energy physics to condensed matter physics.

\section{Vortex solutions and stability} 

The class of non-Hermitian models that we consider have the following equations of motion for the classical, $c$-number complex scalar fields $\phi_1$ and $\phi_2$:
\begin{subequations}
    \label{eq:EoMs}
\begin{align}
    \partial_\mu\partial^\mu\phi_1 + m_1^2\phi_1 + m_3^2\phi_2 + 2\lambda_1|\phi_1|^2\phi_1 &= 0 \,, \\
    \partial_\mu\partial^\mu\phi_2 + m_2^2\phi_2 - m_3^2\phi_1 + 2\lambda_2|\phi_2|^2\phi_2 &= 0 \,,
\end{align}
\end{subequations}
where $m_1^2$, $m_2^2$, $m_3^2$, $\lambda_1$ and $\lambda_2$ are all real parameters. The non-Hermiticity of this system is manifest in the relative sign in the mass mixing terms (proportional to $m_3^2$) between each of the two equations of motion. We work in Minkowski spacetime and use the mostly minus signature convention $(+,-,-,-)$.

The physical viability of this non-Hermitian field theory stems from the symmetry of the underlying Lagrangian formulation (see Eq.~\eqref{eq:LNHtilde} of the appendix) under the combined action of parity $\mathcal{P}$ and time-reversal $\mathcal{T}$ if one of the two complex scalar fields is taken to transform as a pseudo-scalar under parity. There will then be regions of parameter space in which the eigenspectrum is real (regions of unbroken $\mathcal{PT}$ symmetry), regions in which the eigenspectrum features complex conjugate pairs of energies (regions of broken $\mathcal{PT}$ symmetry), and so-called exceptional points, lying at the boundaries between these regions and corresponding to the so-called $\mathcal{PT}$ phase transition. These exceptional points are characterised by eigenvalue coalescence and merger of the corresponding eigenvectors. A primary diagnostic of pseudo-Hermitian dynamics is then the non-linear splitting of eigenvalues as we perturb away from these exceptional points, with the eigenspectrum exhibiting a square-root singularity that cannot occur for a Hermitian theory.

The model described above possesses a global $U(1)$ symmetry. The spontaneous breaking of global and local $U(1)$ symmetries of this model have previously been considered in the context of the Goldstone theorem~\cite{Alexandre:2018uol, Fring:2019hue, Fring:2019xgw} and the Higgs mechanism~\cite{Mannheim:2018dur, Alexandre:2018xyy, Alexandre:2019jdb, Fring:2020bvr}. Additionally, vortices will generically form via the Kibble mechanism~\cite{Kibble:1976sj} and can be identified when a closed loop in space winds an integer number of times around the vacuum manifold (see Refs.~\cite{V&Sbook,M&Sbook} for more detail on cosmic strings and topological defects, in general).

We now proceed to consider vortex solutions by making the familiar Ansatz $\phi_i = f_i(r)e^{in\theta}$, where $n$ is the vortex winding number. The equations of motion~\eqref{eq:EoMs} then take the forms
\begin{align}
    \frac{{\rm d}^2f_1}{{\rm d}r^2} + \frac{1}{r}\frac{{\rm d}f_1}{{\rm d}r} - \frac{n^2}{r^2}f_1 - m_1^2f_1 - m_3^2f_2 - 2\lambda_1f_1^3 &= 0 \,, \nonumber \\
    \frac{{\rm d}^2f_2}{{\rm d}r^2} + \frac{1}{r}\frac{{\rm d}f_2}{{\rm d}r} - \frac{n^2}{r^2}f_2 - m_2^2 f_2 + m_3^2f_1 - 2\lambda_2f_2^3 &= 0 \,.
    \label{eq: string EoMs}
\end{align}

\begin{figure}[!t]
    \centering
    \includegraphics[trim={1.5cm 0.2cm 1.9cm 1.8cm},clip,width=0.48\textwidth]{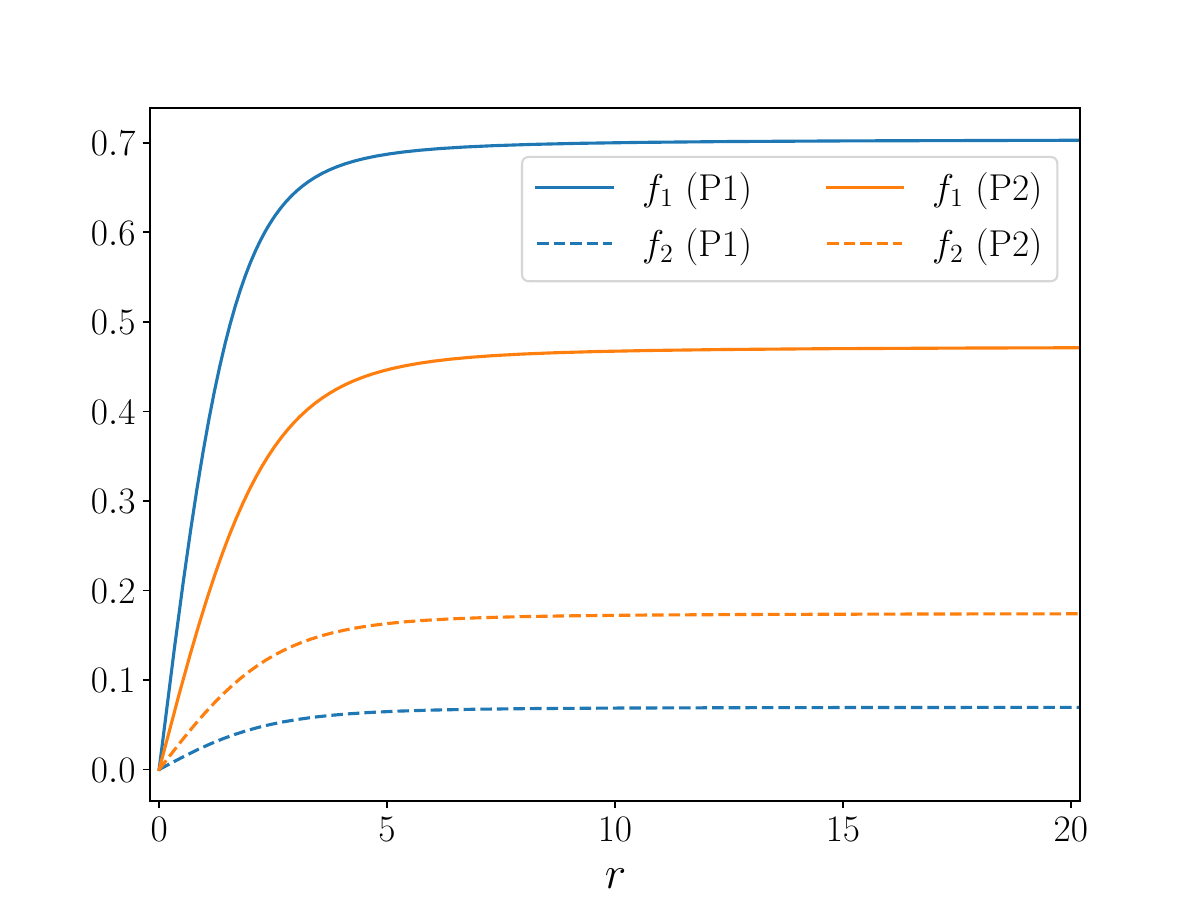}
    \caption{The profiles of string solutions for two different parameter sets. Both cases have $m_1^2=-1$, $\lambda_1=\lambda_2=1$ but parameter set P1 has $m_2^2=1$ and $m_3^2=0.1$, while parameter set P2 has $m_2^2 = 4$ and $m_3^2 = 1.5$. The vortex solutions for parameter set P1 have been presented previously in ref.~\cite{Begun:2022ufc}.}
    \label{fig: string sols}
\end{figure}

We solve these equations numerically on a grid with $10,000$ points and a spacing of $\Delta r = 0.01$ and present the solutions in Figure \ref{fig: string sols}. By rescaling lengths and field magnitudes, we can fix $\lambda_1=\pm1$ and $m_1^2=\pm 1$ without loss of generality,  and we adopt this approach throughout this paper. This sets the length scales of the system to be in units of $|m_1|^{-1}$ and energy densities to be in units of $m_1^4/|\lambda_1|$. We choose to focus on the case where $\lambda_1 = 1$ and $m_1^2 = -1$. Fig.~\ref{fig: string sols} shows two examples of string solutions in different parameter sets: P1 has $m_2^2 = 1$, $m_3^2 = 0.1$ and $\lambda_2=1$, while P2 has $m_2^2=4$, $m_3^2 = 1.5$ and $\lambda_2 = 1$.

In the same way that we can understand the stability of the vacuum states by calculating the eigenvalues of the fluctuation matrix~\cite{Begun:2022ufc}, we can also analyse the stability of the string solution by looking at the linearised equations of motion for perturbations about the string solution, $\phi_i = (f_i+\delta f_i)e^{in\theta}$, where $f_i$ are the solutions to Eq.~\eqref{eq: string EoMs}. Taking the Fourier transform of the perturbation,
\begin{equation}
\delta f_i(t,r,\theta,z) = \sum\limits_{m\in\mathbb{Z}}\int{\rm d}\nu\, {\rm d}k\;\hat{\delta f}_i e^{i(\nu t + m\theta + kz)}\,,
\end{equation}
we obtain the eigenvalue problem
\begin{subequations}
\label{eq:linearised}
\begin{align}
    &\left[ -\frac{{\rm d}^2}{{\rm d} r^2} - \frac{1}{r}\frac{{\rm d}}{{\rm d}r} + \frac{n^2+m^2}{r^2} + m_+^2 + M \right]\mathbf{v} = \Lambda \mathbf{v}\,, \\
    &M = \begin{pmatrix}
        2\beta_1 + \bar{m}^2 & \beta_1 & m_3^2 & 0 \\
        \beta_1 & 2\beta_1 + \bar{m}^2 & 0 & m_3^2 \\
        -m_3^2 & 0 & 2\beta_2 - \bar{m}^2 & \beta_2 \\
        0 & -m_3^2 & \beta_2 & 2\beta_2 - \bar{m}^2
    \end{pmatrix} \nonumber \\
    &\quad\, + \begin{pmatrix}
        \frac{2nm}{r^2} & 0 & 0 & 0 \\
        0 & -\frac{2nm}{r^2} & 0 & 0 \\
        0 & 0 & \frac{2nm}{r^2} & 0 \\
        0 & 0 & 0 & -\frac{2nm}{r^2}
    \end{pmatrix} \,.
\end{align}
\end{subequations}
Herein, we have defined $\Lambda = \nu^2 - k^2$, $\mathbf{v}^T = \begin{pmatrix} \hat{\delta f_1} & \tilde{\delta f_1^*} & \hat{\delta f_2} & \tilde{\delta f_2^*} \end{pmatrix}$ with $\Tilde{\delta f}_i^* = \hat{\delta f_i^*}(-\nu^*, r, -m, -k^*)$, as well as the shorthand notations, $\beta_a = 2\lambda_af_a^2$, $m_+^2 = \frac{1}{2}(m_1^2 + m_2^2)$ and $\bar{m}^2 = \frac{1}{2}(m_1^2-m_2^2)$. The summation and integration over $\nu$, $m$ and $k$ can be removed because each mode must independently satisfy these equations if it is to be a solution for all $t$, $\theta$ and $z$.

The equality $\Lambda = \nu^2 - k^2$ allows for multiple solutions corresponding to each eigenvalue, however, only those that can be boosted into a frame where $k$ is real are physical solutions, otherwise there is no way to excite such a perturbation. In this special frame, it is clear that the usual criterion for stability applies:~only real and positive eigenvalues correspond to stable perturbations. All other eigenvalues must have a complex $\nu$, and the fastest growing mode will be observed in the same special frame. We can, therefore, only say that a string is completely stable if, for every $m$, the eigenvalues are all real and positive. 

While the underlying theory is $\mathcal{PT}$-symmetric, the presence of the string solution has the potential to break this $\mathcal{PT}$-symmetry, such that the spectrum may contain complex-conjugate pairs of eigenvalues, signalling an instability of the string solution. To identify whether this is the case, we first note that the equations of motion for the fluctuations essentially contain spacetime-dependent parameters, due to the spatial variation of the background solution. It has previously been observed~\cite{Chernodub:2021waz, Chernodub:2024lkr} that otherwise $\mathcal{PT}$-symmetric theories with local Lagrangian parameters exhibit mode-specific exceptional points, such that certain fluctuations reside in the $\mathcal{PT}$-broken regime while others reside in the $\mathcal{PT}$-unbroken regime. We will find evidence of the same effect for the present system, identifying a dominant imaginary eigenvalue that drives an instability in the system.

We begin by attempting to diagonalize the eigensystem for the fluctuations via a similarity transformation.  Since the parameters depend upon distance from the string, this similarity transformation will be local and, as a result, it will not commute with the radial derivatives. While this complicates the analysis, it suffices for us to identify any branch cuts in the diagonal (or indeed symmetric) representation of this system.  To this end, we make the similarity transformation, $\mathbf{v} = S\tilde{\mathbf{v}}$, so that the eigenvalue problem becomes
%
\begin{align}
    \label{eq: sim-linearised}
    &\bigg[ -\frac{{\rm d}^2}{{\rm d} r^2} - \frac{1}{r}\frac{{\rm d}}{{\rm d}r} - 2S^{-1}S^\prime\frac{{\rm d}}{{\rm d}r} - S^{-1}S^{\prime\prime} - \frac{S^{-1}}{r}S^\prime \nonumber \\
    &+ \frac{n^2+m^2}{r^2} + \frac{1}{2}(m_1^2+m_2^2) + S^{-1}M S\bigg]\tilde{\mathbf{v}} = \Lambda \tilde{\mathbf{v}}\,,
\end{align}
%
where $S^\prime = \frac{{\rm d}S}{{\rm d}r}$. For the case of $m=0$, we can diagonalise it by setting
%
\begin{align}
    S = 2m_3^2\begin{pmatrix}
        1 & -G_1 & -1 & G_2 \\
        -1 & G_1 & -1 & G_2 \\
        -G_1 & 1 & G_2 & -1 \\
        G_1 & -1 & G_2 & -1 
    \end{pmatrix} \,.
\end{align}
%
Herein, 
\begin{equation}
    G_a = F_a + {\rm sgn}(m_3^2)\sqrt{F_a^2-1}\;,
\end{equation}
where
\begin{subequations}
\begin{align}
F_1(r) &= (m_1^2 - m_2^2 + 2\lambda_1f_1^2 - 2\lambda_2f_2^2)/2m_3^2\,,\\
F_2(r) &= (m_1^2 - m_2^2 + 6\lambda_1f_1^2 -6\lambda_2f_2^2)/2m_3^2\;,
\end{align}
\end{subequations}
and the eigenvalues are
\begin{equation}
\label{eq:eigenvalue_singularity}
\frac{1}{2}(\beta_1+\beta_2) \pm m_3^2\sqrt{F_a^2-1}\;.
\end{equation}
The similarity transformation to diagonalise $M$ is not unique and we have used some of the remaining freedom to select a specific form that produces symmetric matrices $S^{-1}S^\prime$ and $S^{-1}S^{\prime\prime}$. In this way, Eqs.~\eqref{eq: sim-linearised}--\eqref{eq:eigenvalue_singularity} have been expressed in a form that makes it clear that the antilinear symmetry is locally unbroken if both $F_1^2>1$ and $F_2^2>1$, and locally broken otherwise. Importantly, we see the signature of the square-root singularities that evidence the pseudo-Hermitian nature of this theory, occurring here as we approach the exceptional points at $F_a=1$.

Having identified the location of the mode-dependent $\mathcal{PT}$ phase transition, and using just the string solutions in Fig. \ref{fig: string sols}, we therefore do not expect to see an instability in parameter set P2, where $F_1^2$ and $F_2^2$ are greater than $1$ for all $r$, whereas for P1, $F_2$ is negative at the core of the string and positive in the vacuum, so an instability is likely, although not necessarily guaranteed, as gradients still need to be taken into account. 

In order to check this intuition, we discretise Eq.~\eqref{eq:linearised}, so that we can numerically calculate the eigenvalues for this system. For both string solutions, we find many eigenvalues with small spurious imaginary components compared to their real parts. We emphasise the importance of convergence testing here, which allows us to distinguish eigenvalues corresponding to genuine instabilities from the rest, see the appendix for further details. For the solutions in parameter set P1, we indeed find a converged complex eigenvalue when $m=0$, suggesting that a growing mode exists with $\Lambda = 1.592-0.022i$, whereas for parameter set P2, we do not find evidence of any growing modes.  This is in complete agreement with our expectations from the similarity transformation. Moreover, this is confirmation that the system cannot be mapped onto a Hermitian theory, since the eigenvalue is complex. As a result, the instability will have a qualitatively different behaviour:\ it will both grow \textit{and} oscillate, whereas eigenfunctions corresponding to real eigenvalues will either grow/decay \textit{or} oscillate.

\begin{figure}[!t]
    \centering
    \subfloat[Unstable mode.]{
        \centering
        \includegraphics[trim={1.5cm 0.2cm 1.5cm 1.6cm},clip,width=0.48\textwidth]{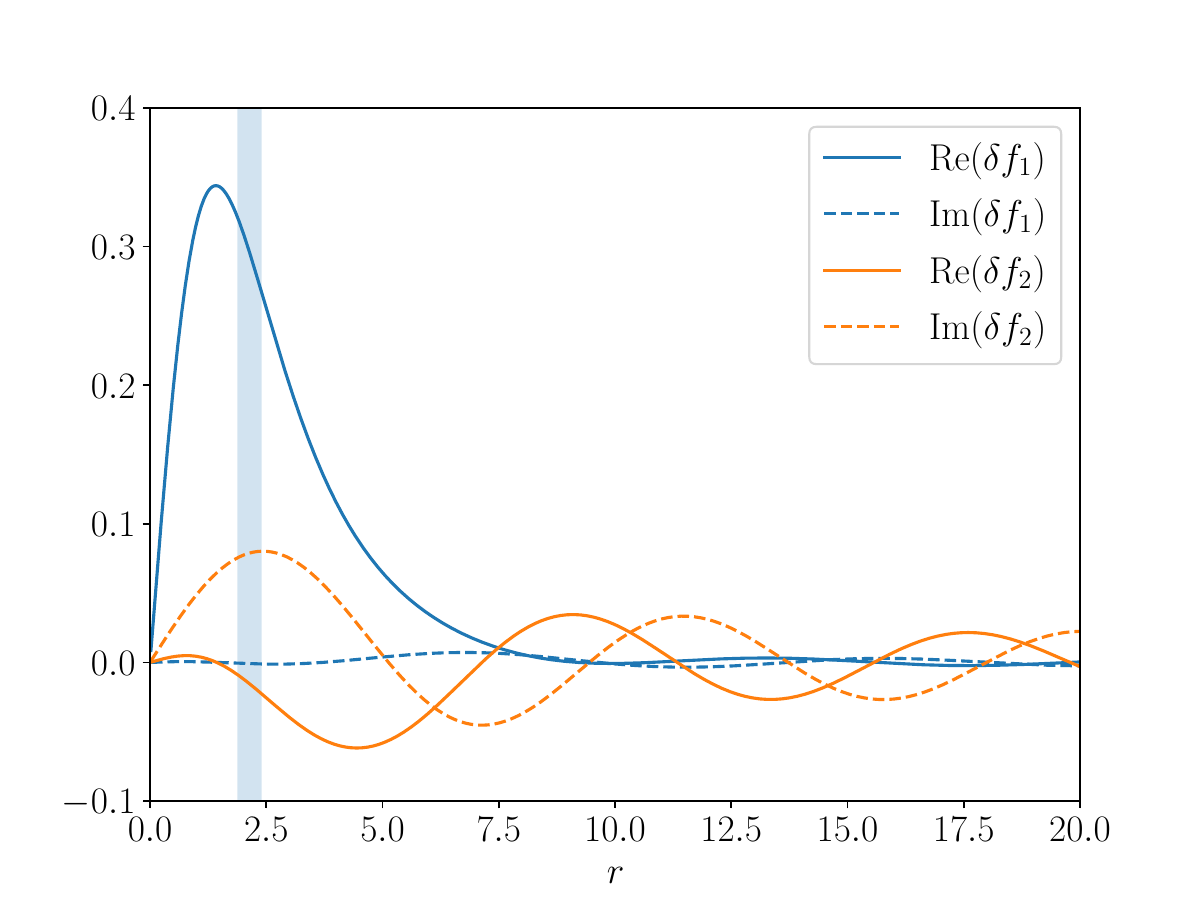}
    }\hfill
    \subfloat[Stable mode.]{
        \centering
        \includegraphics[trim={1.5cm 0.2cm 1.5cm 1.6cm},clip,width=0.48\textwidth]{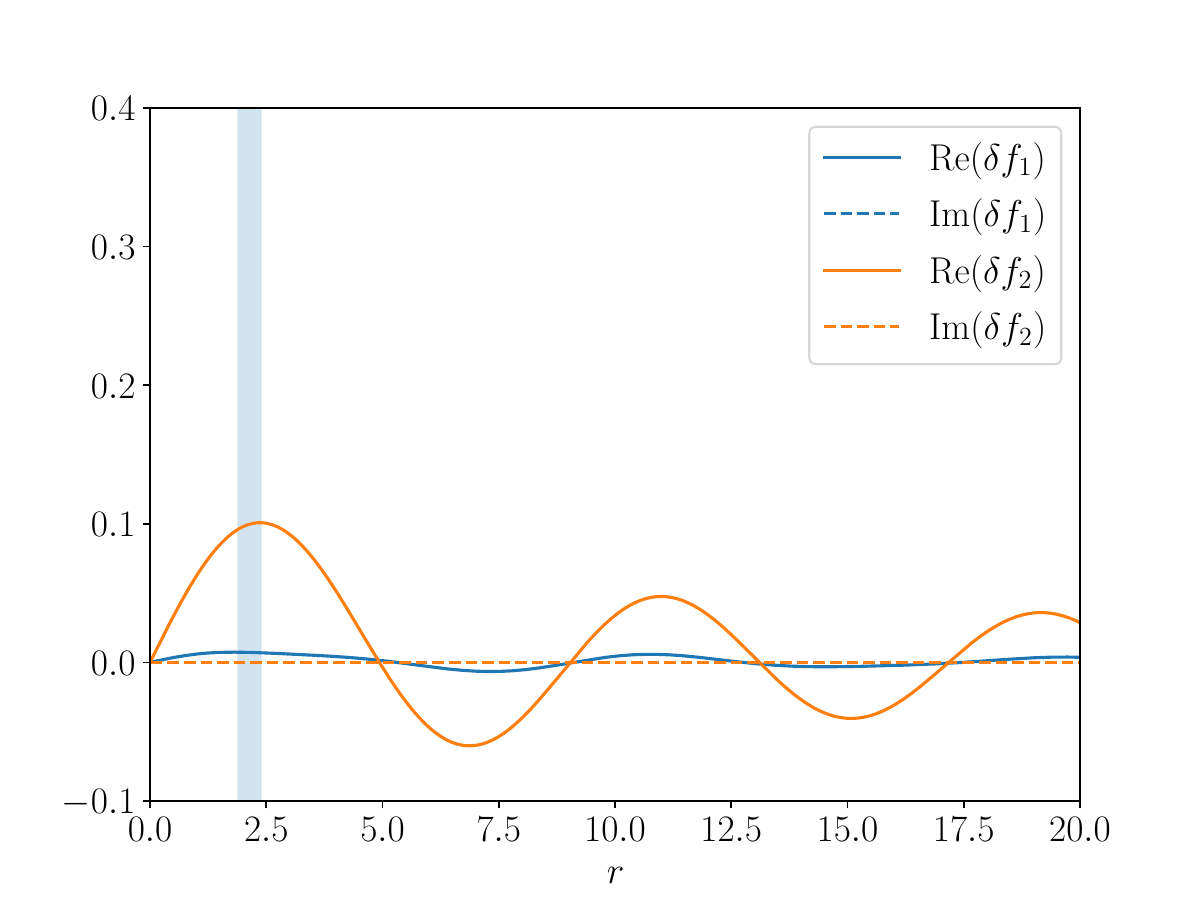}
    }\hfill
    \caption{Two of the eigenfunctions for parameter set P1 with $m=0$, normalised so that $\int r{\rm d}r \mathbf{v}^\dagger\mathbf{v} = 1$. One (a) corresponds to an instability with eigenvalue $\Lambda = 1.592-0.022i$ and the other (b) has the nearby eigenvalue of $\Lambda = 1.601$. As $m=0$, the $\tilde{\delta f_a}^*$ functions are identical to the ones presented here, so we have omitted them from the plots. The shaded band corresponds to the narrow region within which the linearised system can be described as locally non-Hermitian and it is clear that the unstable mode is concentrated around this zone. In contrast, the stable mode is more spread out despite having a similar typical oscillation length scale.}
    \label{fig: eigenfunctions}
\end{figure}

In Fig.~\ref{fig: eigenfunctions}, we present two eigenfunctions for parameter set P1. The shaded region represents the range of $r$ for which $F_2^2<1$. We see that the unstable mode with $\Lambda = 1.592-0.022i$, shown in Fig.~\ref{fig: eigenfunctions}(a), has significantly stronger support and is more localised in this region than the stable mode with the nearby eigenvalue $\Lambda = 1.601$, shown in Fig.~\ref{fig: eigenfunctions}(b). Nevertheless, there is only a narrow region in which the antilinear symmetry is broken locally, which likely explains why the imaginary component of the eigenvalue is small compared to the real component.

\section{Numerical simulations} 

We can directly test the predictions that we made in the previous section by performing numerical simulations of the string solutions. We choose to do these simulations in two spatial dimensions on a Cartesian grid with $1,024$ points and a grid spacing of $\Delta x = 0.125$ in both directions, and evolve forward in time with steps of $\Delta t = 0.025$. Our initial conditions are set by interpolating our solutions for $\phi_a(r,\theta)$ onto the Cartesian grid with zero initial velocity, and we use Dirichlet boundary conditions at the edges of our simulation box. In principle, we could add some additional perturbations on top of these solutions in order to excite specific modes, but here we simply rely on the errors associated with the discretization to source the initial perturbations. If we have successfully identified the fastest-growing unstable modes, then these should come to dominate the motion of the fields, allowing us to make clear comparisons to our analytic expectations.

\begin{figure}[!t]
    \centering
    \subfloat[Evidence of an instability in parameter set P1.]{
        \centering
        \includegraphics[trim={0cm 0.8cm 0.6cm 0.6cm},clip,width=0.48\textwidth]{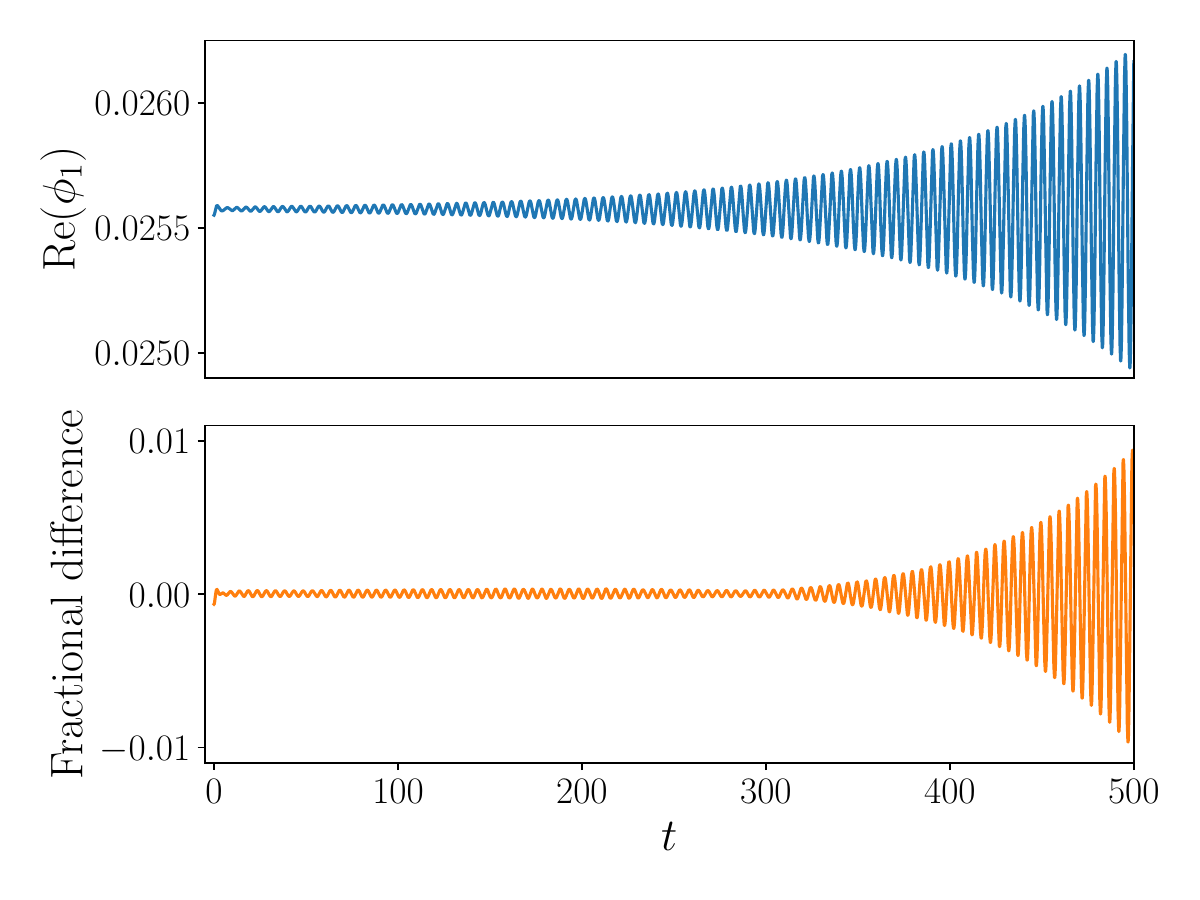}
        \label{fig: field oscillations a}
    }\hfill
    \subfloat[No evidence of instability in parameter set P2.]{
        \centering
        \includegraphics[trim={0.7cm 0.8cm 0.5cm 0.5cm},clip,width=0.48\textwidth]{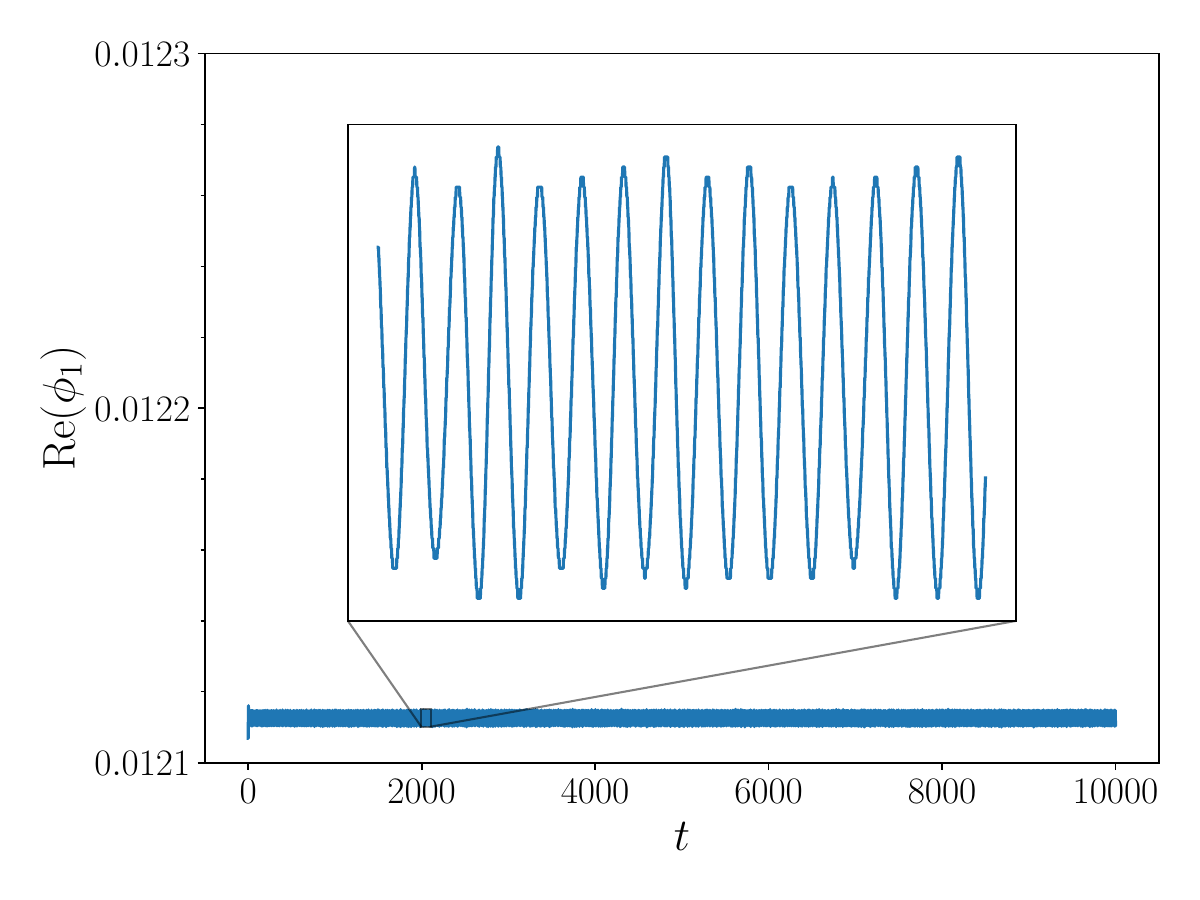}
    }\hfill
    \caption{Evolution of Re$(\phi_1)$ at $x = y = \frac{1}{2}\Delta x$ relative to the core of the string in both parameter sets. In the lower panel of (a), we also show the fractional difference between our prediction from the stability analysis and the results shown in the top panel. The P1 string is clearly unstable and matches the expectation from our stability analysis very well, albeit with a growing difference developing between the two. This difference only reaches a maximum of $0.01$ by $t=500$, which is about $8$ light-crossing times. In contrast, the P2 string appears stable, even when tested over very long timescales as presented here. Note that the other components of the fields have very similar behaviour.}
    \label{fig: field oscillations}
\end{figure}

The core of the string solutions are in the centre of the grid, which does not lie on a lattice site. In Fig.~\ref{fig: field oscillations}, we use the real component of $\phi_1$ at one of the four equidistant lattice sites (in this case at $x=\Delta x/2$ and $y=\Delta y/2$ with respect to the string position) to track the evolution and note that the other components all behave very similarly. As expected, the string in the parameter set P1 exhibits an instability that is very well described by a function of the form $\bar{\phi} + A_i\cos\left({\rm Re}(\nu)t + \alpha\right)e^{-{\rm Im}(\nu)t}$ --- the fractional difference between this and the simulation results can be seen in the lower panel of Fig.~\ref{fig: field oscillations a}. We use $\nu = 1.26-8.72\times 10^{-3}i$ set by the square root of the eigenvalue of the growing mode, an initial offset $\bar{\phi}$ set by the average field value, as well as an initial amplitude $A_i = 8\times 10^{-6}$ and a phase of $\alpha = 1.2\pi$, which were chosen by eye. In contrast, the string in parameter set P2 does not display any signs of an instability, even over much longer time scales. The most obvious oscillation in this data would be well described by $\nu\sim 1$, and indeed we do find eigenvalues that produce similar values. However, as there is no evidence of a growing oscillation, it is not possible to make an \textit{a priori} claim for which frequency will be dominant in the signal.

\begin{figure}[!t]
    \centering
    \subfloat[$t=0$.]{
        \centering
        \includegraphics[trim={0.8cm 0.8cm 1.6cm 0.7cm},clip,width=0.48\textwidth]{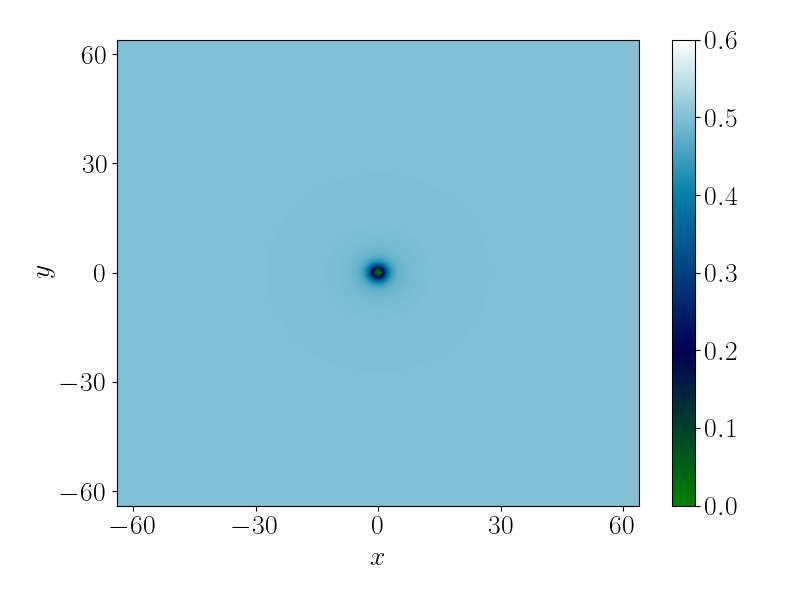}
    }\hfill
    \subfloat[$t=1000$.]{
        \centering
        \includegraphics[trim={0.8cm 0.8cm 1.6cm 0.7cm},clip,width=0.48\textwidth]{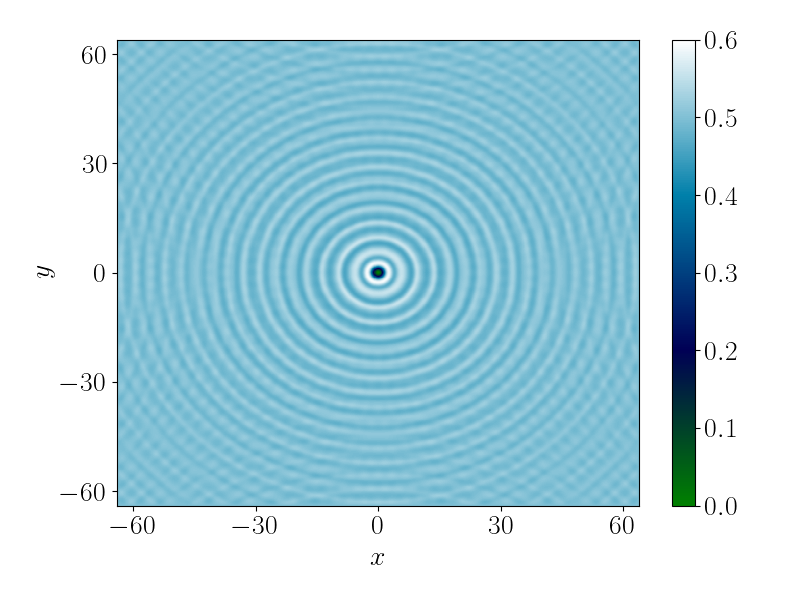}
    }\hfill
    \caption{Snapshots of the total field magnitude, $|\phi_1|^2 + |\phi_2|^2$, during 2D simulations in parameter set P1. The upper plot shows the initial conditions of the simulation, and the lower shows the state of the system by $t=1000$. No perturbations have been added to the initial state; the unstable mode is being sourced purely by imperfections in the solution. It is clear that the dominant instability is an axially symmetric mode, with an apparent wavelength $\sim 4$. Some interference patterns are visible towards the edges of the simulation, but they are small and unlikely to be the cause of the instability because of the circular symmetry of the pattern, as opposed to the rectangular symmetry that would be expected from grid effects.}
    \label{fig: P1 2D heatmap}
\end{figure}

Let us now examine the two cases in a little more detail. In Fig.~\ref{fig: P1 2D heatmap}, we show a heat map of the combined magnitude, $|\phi_1|^2 + |\phi_2|^2$, at two snapshots of the simulation in parameter set P1. At the core of the string, this quantity goes to zero, which is the only apparent feature of the initial conditions. At a much later timestep, $t=1000$, the unstable eigenmode has grown sufficiently to be clearly visible, and it is evident that it primarily corresponds to an axially symmetric mode, as predicted. There is also a characteristic length scale of roughly $4$, which is also broadly consistent with the eigenmode that we have identified. Some unphysical effects are introduced due to interactions with the boundary conditions, since the timescale for these effects to become visible is an order of magnitude larger than the light crossing time but, visually, these still appear to be subdominant. We also note that, by $t\approx 3000$, the instability appears to have completely destroyed the string but, since the effects of the boundaries are also much more obvious, we are uncertain that this is the ``true" evolution and mention this only as an indication of the ultimate fate of the string, which should be viewed with some scepticism.

\begin{figure}[!t]
    \centering
    \includegraphics[trim={0.2cm 0.2cm 2cm 1.8cm},clip,width=0.48\textwidth]{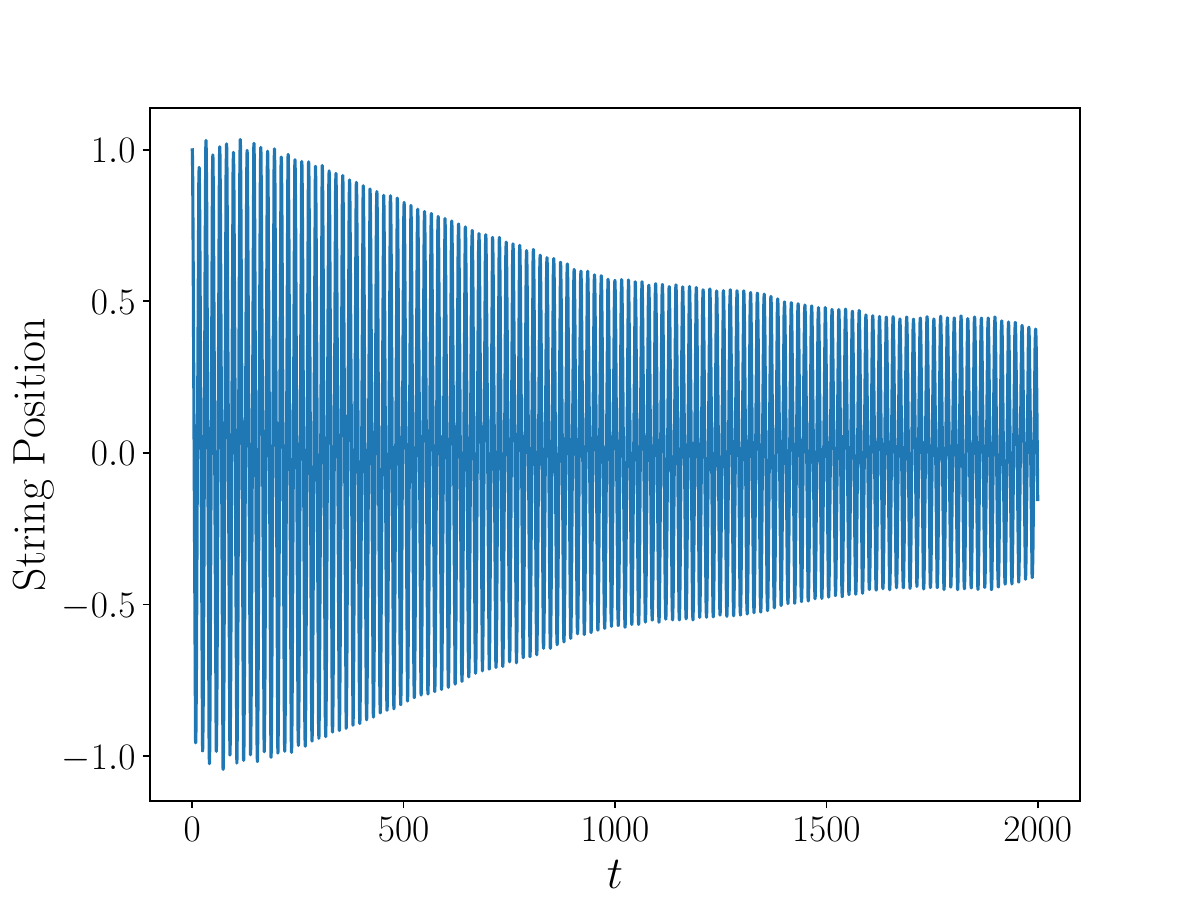}
    \caption{We show how the position of the P2 string along the $x=z=0$ line changes throughout a 3D simulation where we have set up the initial conditions to represent the string solution but sinusoidally displaced as a function of $z$. Even in this scenario with a manually applied perturbation, the string shows no signs of instability.}
    \label{fig: 3D oscillating string P2}
\end{figure}

As for the string in parameter set P2, since it appeared to be stable, we have additionally simulated it in three dimensions as a further test. In this simulation, we use $2,048$ grid points in the $x$ and $y$ directions and Dirichlet boundary conditions, whereas in the $z$ direction we use $64$ grid points and periodic boundary conditions, with a lattice spacing of $0.25$ in all directions and $\Delta t = 0.05$. We perturb the $y$ position of the string sinusoidally, as a function of $z$, so that each $z$ slice is initialised with the 2D string solution but offset slightly from each other. The oscillation of the string will emit radiation that will reflect off the boundaries and back-react on the string, as the light crossing time across constant $z$ planes is only $t=256$. It is possible that this could suppress any growing modes, but we run the simulation past the light crossing time regardless in order to have a wider dynamic range, since we expect the timescale of any instabilities to be large. We track the position of the string along the $x=z=0$ line by detecting the position where the imaginary component of $\phi_1$ changes sign\footnote{We also track the location where the imaginary component of $\phi_2$ changes sign, but it is indistinguishable by eye for this simulation.}, which is presented in Fig.~\ref{fig: 3D oscillating string P2}. Again, we see no signs of any instability in the evolution of the string, even in this case with an applied perturbation.

\section{Conclusion} 

In this {\em letter}, we have examined analytical expectations for the dynamics of vortices in a non-Hermitian field theory and compared these to the results of numerical simulations. By considering two sets of sample parameters, we have presented two vortex solutions, one of which appears to be stable, even on long timescales, and one of which is demonstrated to be unstable. A stability analysis has also been performed previously for solitons with analytical solutions in the non-Hermitian extension of the Bullough--Dodd model~\cite{CORREA2022115783}, also finding two classes of solutions: one stable and one unstable. For the model considered in this work, we have shown that the instability can be attributed to local breaking of the antilinear symmetry.

It is reasonable to expect that stable vortices of non-Hermitian theories could behave in a manner comparable with those of Hermitian theories, but it is the unstable variety that are novel and more interesting from a phenomenological perspective. For example, in the case of cosmology, the standard lore is that the evolution of domain walls formed after a discrete symmetry breaking and monopoles formed at the GUT transition would both come to dominate the universe. These are known as the domain wall and monopole problems. Avoiding these heavy relics was one of the original motivations for inflation. If topological defects can be intrinsically unstable in non-Hermitian theories then such relics could decay during the evolution of the Universe. This would mean that discrete symmetries and those which create monopoles may not necessarily be a problem when realised within pseudo-Hermitian theories. 

However, it is yet to be seen whether these instabilities ultimately lead to a resolution for these defect domination scenarios, or some other complicated dynamical process, such as unbounded growth of the instability (see the discussion of Eq.\eqref{eq: conserved energy-momentum} in the appendix). While this might be considered pathological in Minkowski spacetime, such an instability may be desirable in curved spacetimes and for cosmological applications. An example has been considered in Ref.~\cite{Copeland:2025qiu}, where the interplay between $\mathcal{PT}$-symmetry breaking and Hubble damping for the same non-Hermitian theory studied here can give rise to a novel mechanism for driving inflation or late-time dark energy. These are intriguing possibilities which we intend to further explore in an upcoming work by modelling the formation and evolution of non-Hermitian vortices following a phase transition.

\section*{Acknowledgements}

    The authors thank Maxim Chernodub and Carl M. Bender for helpful discussions and would like to acknowledge the assistance given by Research IT and the use of the Computational Shared Facility at The University of Manchester. This work was supported by the Science and Technology Facilities Council (STFC) [Grant No.~ST/X00077X/1]; a UKRI Future Leaders Fellowship [Grant No.~MR/V021974/2] the International Exchanges project of the Royal Society No.~IES\textbackslash R3\textbackslash203069; and by the International Emerging Actions (IEA) project ``Non-Hermitian field theories for particle and solid-state physics / TH\'ECHNO'' of CNRS.

\appendix

\section{Lagrangian formulations}

The equations of motion can be obtained from a family of Lagrangians with the general form
\begin{align}
    \label{eq:LNHtilde}
    \mathcal{L}_{\text{NH}} &= \partial_\mu\tilde{\phi}_1^*\partial^\mu\phi_1 + \partial_\mu\tilde{\phi}_2^*\partial^\mu\phi_2 - m_1^2\tilde{\phi}_1^*\phi_1 - m_2^2\tilde{\phi}^*_2\phi_2 \nonumber \\
    &- m_3^2(\tilde{\phi}_1^*\phi_2 - \tilde{\phi}_2^*\phi_1) - \lambda_1(\tilde{\phi}_1^*\phi_1)^2 - \lambda_2(\tilde{\phi}_2^*\phi_2)^2 \,,
\end{align}
wherein $\tilde{\phi}_1^*$ and $\tilde{\phi}^*_2$ are the respective dual fields of $\phi_1$ and $\phi_2$. The presence of the tilde is to emphasise that, due to the non-Hermiticity of the Lagrangian, the dual fields are, in general, not simply the complex conjugates~\cite{Alexandre:2020gah, Alexandre:2022uns, Sablevice:2023odu}. Similar observations have been made in the context of second-order fermionic theories~\cite{LeClair:2007iy, Ferro-Hernandez:2023ymz}. This is easily justified when realising that the field $\phi$ evolves with the Hamiltonian $H$, while the complex-conjugate field $\phi^*$ evolves with the Hamiltonian $H^*\neq H$~\cite{Chernodub:2021waz}. More generally, composite operators such as $\phi_i^*\phi_j$ do not transform properly under the generators of the Poincaré group~\cite{Sablevice:2023odu}.

As described earlier, the physical viability of this non-Hermitian field theory stems from the symmetry of the Lagrangian~\eqref{eq:LNHtilde} under the antilinear transformation $\phi_1 \to \tilde{\phi}_1^*$ and $\phi_2 \to -\tilde{\phi}_2^*$. These transformations of the $c$-number fields can be realised as the combined action of parity $\mathcal{P}$ and time-reversal $\mathcal{T}$ if one of the two complex scalar fields is taken to transform as a pseudo-scalar.  More specifically, there exists a parity like transformation $\eta_{\mathcal{P}}$ such that, e.g., $\eta_{\mathcal{P}}: \phi_2(x)\mapsto \phi_2^{\eta_{\mathcal{P}}}(x_{\mathcal{P}})=-\tilde{\phi}_2(x)$ (see Ref.~\cite{Chernodub:2025wga}).  

Since we work entirely at the classical level, there are a number of ways to obtain the equations of motion listed in Eq.~\eqref{eq:EoMs}:
\begin{itemize}
    \item We can force $\tilde{\phi}_i^*=\phi_i^*$ and vary with respect to $\phi_i^*$ only. Note, however, that the resulting Lagrangian
    \begin{align}
    \label{eq:NH1}
    \mathcal{L}_{\text{NH1}} &= \partial_\mu\phi_1^*\partial^\mu\phi_1 + \partial_\mu\phi_2^*\partial^\mu\phi_2 - m_1^2|\phi_1|^2 \nonumber \\
    &- m_2^2|\phi_2|^2 - m_3^2(\phi_1^*\phi_2 - \phi_2^*\phi_1) \nonumber \\
    &- \lambda_1|\phi_1|^4 - \lambda_2|\phi_2|^4
    \end{align}
    will not transform properly under the Poincaré group. Moreover, varying with respect to $\phi_i$ would effect $m_3^2\to -m_3^2$ in the equations of motion, since $\mathcal{L}\neq \mathcal{L}^*$. This change of sign, however, is not physically relevant, as it amounts only to the transformation $\phi_2\to -\phi_2$ (or, equivalently, $\phi_1\to -\phi_1$). Nevertheless, this means that the Euler--Lagrange equations obtained by varying with respect to $\phi_i^*$ and $\phi_i$ are not mutually consistent, except for trivial solutions, such that we must proceed by prescription by fixing the dynamics with respect to a self-consistent subset of the Euler--Lagrange equations~\cite{PhysRevD.96.065027}, viz., those obtained by varying with respect to $\phi_i^*$ only.
    \item We can consider the following similar theory with Hermitian Lagrangian
    \begin{align}
    \mathcal{L}_{\text{NH2}} &= \partial_\mu\phi_1^*\partial^\mu\phi_1 - \partial_\mu\phi_2^*\partial^\mu\phi_2 - m_1^2|\phi_1|^2 \nonumber \\
    &+ m_2^2|\phi_2|^2 - m_3^2(\phi_1^*\phi_2 + \phi_2^*\phi_1) \nonumber \\
    &- \lambda_1|\phi_1|^4 + \lambda_2|\phi_2|^4 \,.
    \label{eq: H2 Lagrangian}
    \end{align}
    This is related to $\mathcal{L}_{\text{NH1}}$ via a similarity transformation, which can be obtained perturbatively and takes the following form:
    \begin{subequations}
    \begin{align}
        \Phi(x)&\to A\cdot \Phi(x)\nonumber\\&\phantom{\to}+i\int\!{\rm d}^4y\;G(x,y)\cdot B(y)+\mathcal{O}(\lambda_i^2)\;,\\
        \Tilde{\Phi}^*(x)&\to \Phi^*(x)\cdot A\nonumber\\&\phantom{\to}+i\int\!{\rm d}^4y\;B^*(y)\cdot G(y,x)+\mathcal{O}(\lambda_i^2)\;,
    \end{align}
    \end{subequations}
    where $\Phi=(\phi_1,\phi_2)$, $G(x,y)$ is the Green's function of the matrix-valued Klein-Gordon operator $\Box+M^2$, with $(\Box+M^2)G(x,y)=-\delta^4(x-y)$, and
    \begin{equation}
        A=\begin{pmatrix}1 & 0 \\ 0 & -i\end{pmatrix}\;,\qquad B=\begin{pmatrix}0 \\ \lambda_2\phi_2^{*2}(y)\phi_2(y)\end{pmatrix}\;,
    \end{equation}
    cf.~the limit $\lambda_2\to 0$ in Ref.~\cite{Mannheim:2018dur}. Note that the Hermiticity of the Lagrangian in Eq.~\eqref{eq: H2 Lagrangian} does not imply that the theory is Hermitian. The latter is manifest in the wrong sign of the $\phi_2$ kinetic term, which indicates that the construction of the state space remains non-trivial.
\end{itemize}

\section{Numerical tests} 

When calculating the eigenvalues of Eq.~\eqref{eq:linearised} numerically, it is important to check that they (or at least the eigenvalues of interest) have sufficiently converged. This is always a crucial step in any numerical study, but it is of particular importance here as we have found many eigenvalues with spurious imaginary components that converge in a manner consistent with them actually being purely real. To put this in context, for our lowest quality lattice, with $N_r = 4000$ sites and a spacing of $\Delta r = 0.04$, we found approximately 500 out of the total 16,000 eigenvalues had an imaginary component with an absolute value greater than $10^{-4}$ but, after convergence testing, we can only confidently say that one of these is genuinely complex valued.

\begin{figure}[!t]
    \centering
    \includegraphics[trim={0.7cm 0.8cm 0.6cm 0.7cm},clip,width=0.48\textwidth]{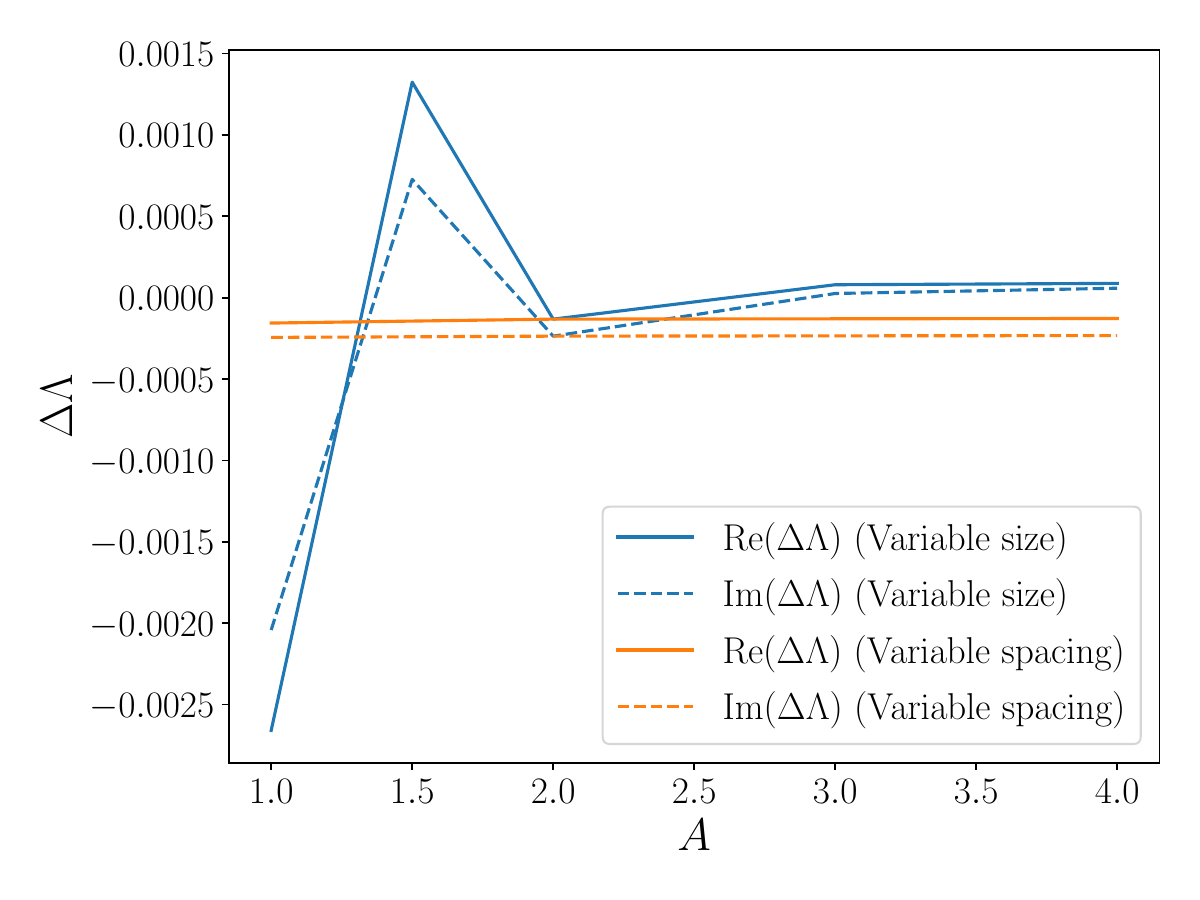}
    \caption{Convergence testing of the complex eigenvalue, $\Lambda = 1.592 - 0.022i$, in parameter set P1, clearly showing that the value has converged to a good degree. For all other eigenvalues of our analysis for this vortex solution, the imaginary component converged to a value that could not confidently be distinguished from zero. We tested changing both the physical size of the grid, while leaving $\Delta r = 0.02$ (variable size), and changing the lattice spacing while keeping $N_r\Delta r = 160$ (variable spacing). We define $A = N_r/4000$ in the former case and $A = 0.04/\Delta r$ in the latter so that the two tests can be displayed on the same plot. }
    \label{fig: convergence}
\end{figure}

In Fig.~\ref{fig: convergence}, we show the convergence of this eigenvalue under variation of both the physical size of the lattice, by varying $N_r$ while keeping $\Delta r=0.02$ fixed, and the lattice spacing, by varying $\Delta r$ while keeping $N_r\Delta r=160$ fixed. To display all the data on the same plot, we have defined $A = N_r/4000$, when varying the size, and $A = 0.04/\Delta r$, when varying the spacing; and we show the difference between the calculated eigenvalue and the value quoted in this paper, $\Delta\Lambda = \Lambda - 1.592+0.022i$. It is clear that the variation of the lattice spacing has a much smaller influence on the numerical value of the eigenvalue compared to the physical size of the lattice. Based on the scale of the variations in the imaginary component of the eigenvalue. We conclude that this eigenvalue has converged to a complex value.

\begin{figure}[!t]
    \centering
    \includegraphics[trim={0.8cm 0.2cm 2cm 1.8cm},clip,width=0.48\textwidth]{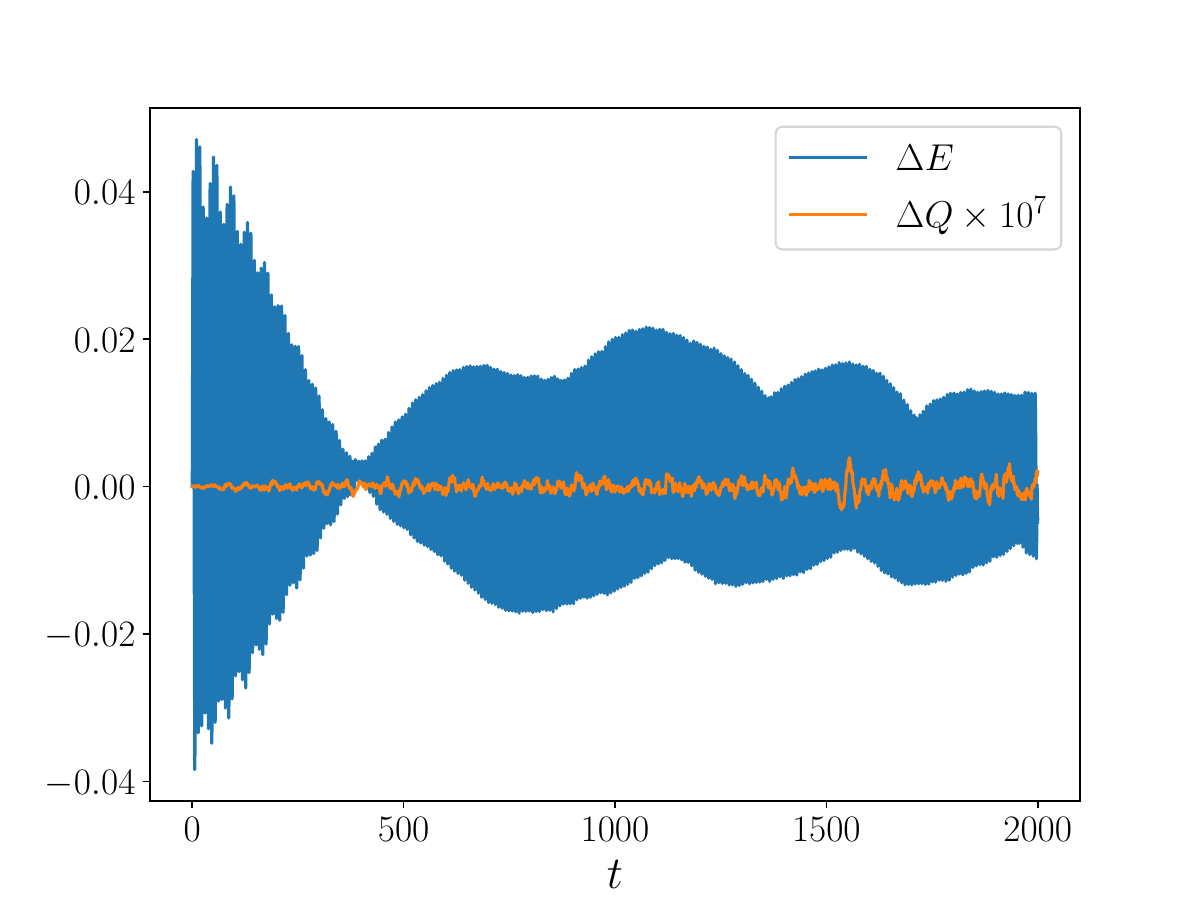}
    \caption{Variation of the total energy and charge, both of which are exactly conserved in the continuum limit, during the simulation of a 3D perturbed string in parameter set P2. The total charge has been scaled up by a factor of $10^7$ so that the variations are visible.}
    \label{fig: conserved quantities}
\end{figure}

We can test the reliability of our simulations by checking that quantities that we expect to be conserved in the continuum limit are conserved to a good degree of accuracy in the discretised system. These are most readily derived from Eq.~\eqref{eq: H2 Lagrangian}, using Noether's theorem and the symmetries of the Lagrangian. The connection of these conserved quantities to the explicitly non-Hermitian Lagrangian in Eq.~\eqref{eq:NH1} is obfuscated by the subtleties of the variational procedure. In the latter case, one finds that Noether's theorem continues to apply in its original formulation, but the conserved current corresponds to a transformation that effects a non-trivial variation of the Lagrangian that cancels against the non-vanishing of a subset of the Euler--Lagrange equations~\cite{PhysRevD.96.065027}. The correspondence between these derivations of conserved $U(1)$ currents was first identified in Ref.~\cite{Mannheim:2018dur}. The conserved current due to the U(1) symmetry is given by
\begin{align}
    j^\mu = i[\phi_1\partial^\mu\phi_1^* - \phi_1^*\partial^\mu\phi_1 - \phi_2\partial^\mu\phi_2^* + \phi_2^*\partial^\mu\phi_2] \,,
\end{align}
and the conserved energy-momentum tensor is
\begin{align}
    T^{\mu\nu} &= \partial^\mu\phi_1^*\partial^\nu\phi_1 + \partial^\nu\phi_1^*\partial^\mu\phi_1 - \partial^\mu\phi_2^*\partial^\nu\phi_2 \nonumber \\
    &- \partial^\nu\phi_2^*\partial^\mu\phi_2 - g^{\mu\nu}\mathcal{L}_{\text{NH2}} \,.
    \label{eq: conserved energy-momentum}
\end{align}
We draw attention to the ghost-like nature of the field $\phi_2$, as it appears in the conserved energy-momentum tensor. This results in a classical energy-momentum tensor that is not bounded from below, in as much as it can be made arbitrarily negative by suitable choice of initial conditions. However, we do not see evidence that the instability observed for P1 is directly attributable to a negative-energy cascade. In particular, the similarity to the instabilities about pure vacuum reported in \cite{Begun:2022ufc} and the apparent stability of P2 suggest that the observed instability of P1 has no direct relation to negative-energy cascades. We instead see evidence that the instability results from the breaking of the $\mathcal{PT}$ symmetry by the spatially varying background of the vortex solution, which drives the dominant unstable mode beyond the exceptional point that marks the $\mathcal{PT}$ phase transition for this theory.

In Fig.~\ref{fig: conserved quantities}, we present the variation of the energy, $E = \int T^{00}{\rm d}^3x$, and the charge, $Q = \int j^0 {\rm d}^3x$, about their values at $t=0$ during the 3D simulation of the oscillating string in parameter set $P2$. The energy varies by $\sim 0.01\%$ as the initial energy is $E(t=0)\approx 500$, where we have normalised the energy so that the vacuum has an energy density of zero. The total charge in the box only varies by $\sim 10^{-10}$, although this is slightly harder to contextualise since all field component velocities are initially zero and therefore so is the total charge. In the 2D simulations, both of these quantities are conserved to an even higher degree of accuracy, including those simulations where a growing instability is present.

\section{Higher angular mode numbers} 
In order to understand the regimes in which $M$ can be considered as locally Hermitian/non-Hermitian for general values of $m$, it is useful to revisit the simpler $m=0$ case in more detail. The similarity matrix that we have used in this simpler case can be written as $S = VD$, where the columns of $V$ are formed from the eigenvectors of $M$, and $D$ is a diagonal matrix that effectively acts as a choice of normalisation for each eigenvector. The basis change enacted by $V$ alone will diagonalise $M$, with $D$ having no effect on the result of $S^{-1}MS$, since it commutes with all diagonal matrices. The reason for its inclusion therefore is to make the $S^{-1}S^\prime$ and $S^{-1}S^{\prime\prime}$ terms symmetric, so that the system preserves the antilinear symmetry if there are no complex numbers, and locally breaks the antilinear symmetry otherwise.

We start by showing that if $S^{-1}S^\prime$ is symmetric, $S^{-1}S^{\prime\prime}$ will also be symmetric. If $S^{-1}S^\prime$ is symmetric, then $S^\prime S^T$ will also be symmetric and vice-versa, as long as the inverse exists. The derivative $S^{\prime\prime}S^T + S^\prime(S^T)^\prime$ is therefore also symmetric, and so is $S^\prime(S^T)^\prime$ individually. This implies that $S^{\prime\prime}S^T$ must be also be symmetric and therefore so is $S^{-1}S^{\prime\prime}$.

Having shown that we only need to concern ourselves with choosing $D$ so that one of the two matrices is symmetric, let us see how it affects the simpler of the two:\ $S^{-1}S^\prime = D^{-1}V^{-1}V^\prime D + D^{-1}D^\prime$. The second term is diagonal and therefore not a concern, while the action of $D$ on the first term is $A_{ab} \to \frac{D_{bb}}{D_{aa}}A_{ab}$. For a general four-by-four matrix, there is an insufficient number of degrees of freedom in $D$ to achieve a symmetric $S^{-1}S^\prime$, which is the reason why the case with $m\neq0$ is technically more involved. Fortunately, in the $m=0$ scenario, the upper right and lower left $2\times 2$ submatrices of $V^{-1}V^\prime$ are zero, so we can always choose $D$ such that $S^{-1}S^\prime$ is a symmetric matrix.

For higher angular mode numbers, the system can still be expressed in a symmetric form by relaxing the requirement that $S^{-1}MS$ is diagonal into a requirement that it is symmetric. We can express a similarity transformation that satisfies this requirement as $S=VDU$, where $U$ is an element of O(4,$\mathbb{C}$), such that $S^{-1}MS = U^TM_\Lambda U$, where $M_\Lambda$ is a diagonalisation of $M$. Restricting to the case where $D=I$, as this extra freedom will turn out to be unnecessary, we would like to have $[V^\prime U + VU^\prime]U^TV^T = VU[(U^T)^\prime V^T + U^T(V^T)^\prime]$. Using $UU^T = I$ and the implied $U^\prime U^T = -U(U^T)^\prime$, we get that $U^\prime = \frac{1}{2}\left[(V^T)^\prime(V^T)^{-1} - V^{-1}V^\prime\right]U$, for which the solution is
%
\begin{align}
    U(r) = \exp\left[\frac{1}{2}\int_0^r\left\{\frac{{\rm d}V^T}{{\rm d}r^\prime}(V^T)^{-1} - V^{-1}\frac{{\rm d}V}{{\rm d}r^\prime}\right\}{\rm d}r^\prime\right]U_0 \,.
\end{align}
%

\bibliographystyle{elsarticle-num}
\bibliography{refs.bib}

\end{document}